\documentclass[aps,prl,floatfix,showpacs,twocolumn,superscriptaddress]{revtex4}
\usepackage{amsmath,amssymb}
\usepackage{graphicx}
\usepackage{epsfig}
\usepackage{psfrag}
\usepackage[usenames]{color}
\usepackage{ulem}

\begin{document}

\hsize\textwidth\columnwidth\hsize\csname@twocolumnfalse\endcsname

\title{Magnetoplasmon Fano resonance in Bose-Fermi mixtures}

\author{M.~Boev}
\affiliation{Institute of Semiconductor Physics, Siberian Branch of Russian Academy of Sciences, Novosibirsk 630090, Russia}

\author{V.~M.~Kovalev}
\email{vadimkovalev@isp.nsc.ru}
\affiliation{Institute of Semiconductor Physics, Siberian Branch of Russian Academy of Sciences, Novosibirsk 630090, Russia}
\affiliation{Department of Applied and Theoretical Physics, Novosibirsk State Technical University, Novosibirsk 630073, Russia}

\author{I.~G.~Savenko}
\affiliation{Center for Theoretical Physics of Complex Systems, Institute for Basic Science, Daejeon, Republic of Korea}
\affiliation{National Research University of Information Technologies, Mechanics and Optics, St. Petersburg 197101, Russia}
\affiliation{Nonlinear Physics Centre, Research School of Physics and Engineering, The Australian National University, Canberra ACT 2601, Australia}


\date{\today}

\begin{abstract}
We investigate theoretically the magnetoplasmon (cyclotron) resonance in a hybrid system consisting of spatially separated two-dimensional layers of electron and dipolar exciton gases coupled via the Coulomb forces. We study the dynamics of this system under the action of weak alternating external electromagnetic  field in the presence of uniform magnetic field, perpendicular to the layers. We reveal that the electromagnetic power absorption exhibits a double-resonance spectrum. We show that the first resonance is associated with the conventional well-studied magnetoplasmon excitations of the electron gas and it has a standard Lorentzian shape, whereas the second resonance is a peculiarity attributed to the Bose-condensed exciton gas. Further, we explicitly demonstrate that the spectrum of the system exhibits an asymmetric Fano-type profile, where the excitonic peak is extremely narrow in comparison with the magnetoplasmon one. We show that the shape of the resonance and the position of the peaks depends on the magnitude of the applied magnetic field, exciton condensate density and exciton-impurity scattering time. In particular, the Fano profile turns into a Lorentzian shape with decreasing exciton-impurity scattering time and the position of the plasmon-associated resonance is mainly sensitive and determined by the magnetic field strength, whereas the exciton-condensate peak position is determined by exciton condensate density. It opens the experimental possibility to determine the latter quantity in cyclotron resonance experiment.



\end{abstract}	


\maketitle


{\it Introduction.---}
Hybrid Bose-Fermi systems represent a testbed for studying many-body phenomena and a promising platform for future applications~\cite{RefImamogluPRB2016}. In cold atomic gases, recent research has been concentrated on study devoted to the Feshbach resonance phenomena~\cite{RefFesh1,RefFesh2,RefFesh3,RefFesh4} and aimed at tuning the strength of two-particle interaction between atoms leading to new types of phase transitions~\cite{RefPhase1,RefPhase2}.
In the solid-state, Bose-Fermi mixtures are usually studied employing exciton or exciton-polariton gases~\cite{RefShelykhPRL2010} which represent Bose-Einstein condensates~\cite{RefButov,RefTimofeev} (BECs) interacting with either electrons and holes coexisting in the same layers or neighbouring two-dimensional layers containing electronic gases (2DEG). In particular, it has been demonstrated that the interaction between a 2DEG and the indirect excitons can lead to the formation of the so-called excitonic supersolid phase~\cite{RefShelykhPRL1051404022010, RefMatuszewskiPRL1080604012012}.

One of the possible physical realisations of Bose-Fermi systems is a semiconductor heterostructure, where the bosonic subsystem is represented by excitons localised in a quantum well (QW) or double quantum well (DQW) and fermions are a 2DEG in a parallel QW. In the lateral (let it be $xy$) plane, both the excitons and electrons can propagate freely. Their motion, self-interaction and exciton-electron interaction result in the possibility for each sort of particles to propagate along different paths which, in order, can interfere between each other.
In the case of the constructive interference, it is reasonable to assume that we can achieve enhancement phenomena and emergence of resonant effects.
Instead, destructive interference might result in the suppression of the particle transport~\cite{RefFano1,RefFano2}.
In this article we are aiming at checking this assumption. In particular, we are interested in the Fano resonance phenomena which is known to be a general type of a resonance and it can be observed in the case of paths interference~\cite{RefMiroshnichenkoFanoReview}.
The Fano resonance is conventionally characterised by a peculiar asymmetric profile of the spectral line. In particular, it usually consists of two peaks and one dip. One of the peaks lies in a close proximity to the dip, manifesting coexistence of resonant transmission and reflection in the system.

\begin{figure}[!t]
	\includegraphics[height=0.35\linewidth]{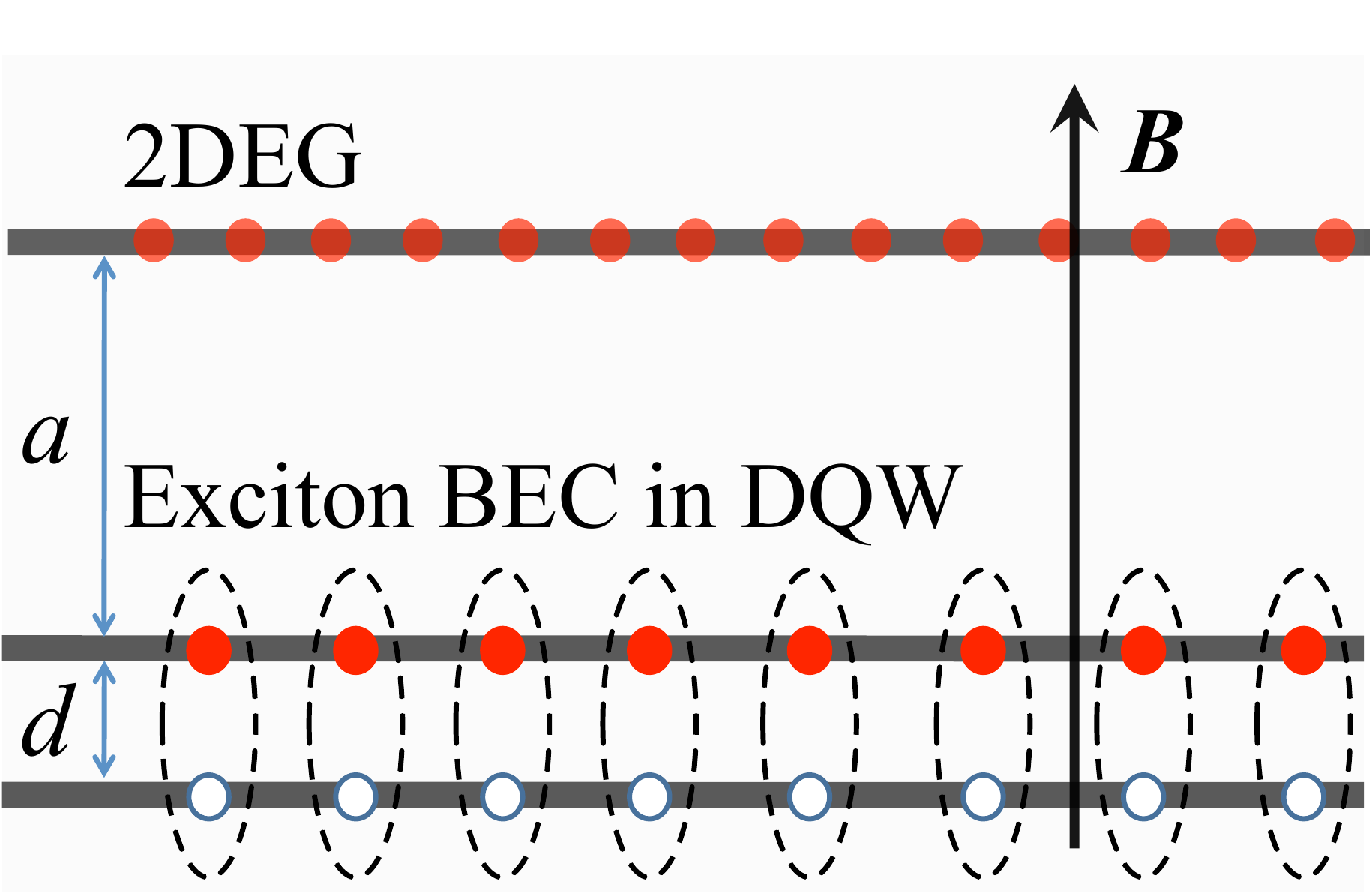}
\caption{System schematic. Excitons are localized in the double quantum well (DQW) structure formed by two  layers containing electrons and holes and separated by distance $d$. Two dimensional electron gas (2DEG) is localized in other QW separated from the DQW by the distance $a$. External applied magnetic field oriented perpendicular to the layers.}
	\label{FigSystem}
\end{figure}

{\it Theory of the magnetoplasmon resonance.---}
We consider a system which represents two parallel layers of electrons and excitons, see Fig.~\ref{FigSystem}. Electrons occupy the QW and form a 2DEG. The excitons are considered to be rigid dipoles, having their dipole moment oriented perpendicular to the plane of the layers containing the DQW and moving freely as a whole within. We assume that the exciton internal degrees of freedom are not excited by the external applied EM field or temperature, thus we consider a zero-temperature case when the quantum effects manifest themselves most clearly.
The uniform magnetic field is directed along the growth axis of the structure,
thus it is oriented perpendicular to the electronic and excitonic layers. We will also assume that the
magnetic field is weak and thus it cannot affect the exciton center-of-mass motion.
In contrast, it can substantially affect the motion of electrons.

Such external EM field, $\textbf{E}(\textbf{r},t)=(E_0,0,0)e^{i\textbf{kr}-i\omega t}$, with the electric field component lying in the plane of the QW, produces deviations of the electronic density from its equilibrium value, $\delta n(\textbf{r},t)=n(\textbf{r},t)-n_0$, where $\textbf{r}$ is a vector coordinate in the QW plane, and $\textbf{k}$ is the in-plane wave vector. In typical experimental structures used to excite 2D plasmons, such a field can be created by laminating a metallic grating above the electron layer.  We assume that EM field does not interact with the excitons directly, nevertheless, the exciton gas density can be influenced by the field via the interaction with electrons, that we will show below.

Dynamics of the electron density can be described by the continuity equation, $\dot{\rho}+\textmd{div} \,\textbf{j}=0$, giving the relation between the Fourier components:
\begin{gather}\label{eq1}
\delta n_{k\omega}=-\frac{k}{e\omega}j_{k\omega},
\end{gather}
where we assume that the vector $\textbf{k}$ is directed along the `$x$' axis. Further we use the Ohm's law,
\begin{gather}\label{eq2}
j_{k\omega}=\sigma_{B}\left(E_0-\frac{1}{e}F_{k\omega}\right),
\end{gather}
where $\sigma_{B}$ is the `$xx$' component of the conductivity tensor in a magnetic field,  $F_{k\omega}$ is the `$x$' component of the applied force, $\textbf{F}(\textbf{r},t)$, which has two contributions: the first one is coming from the electron-electron interaction and the second one is due to the electron-exciton interaction. Using the general relation between a force and the potential energy, $\textbf{F}(\textbf{r},t)=-\nabla W(\textbf{r},t)$,
we find:
\begin{gather}\label{eq3}
F_{k\omega}=-ik(U_{k}\delta n_{k\omega}+V_{k}\delta N_{k\omega}),
\end{gather}
where
\begin{gather}\label{eq4}
U_k=\frac{2\pi e^2}{\epsilon k},\,\,\,V_k=\frac{2\pi e^2}{\epsilon k}\left(1-e^{-kd}\right)e^{-ka}
\end{gather}
are well-known electron-electron and electron-exciton interaction energies, respectively, $\epsilon$ is the dielectric constant. The fluctuating exciton density in the excitonic BEC, $\delta N_{k\omega}$, can be found from the dynamical equations.

The elementary excitations of the Bose-condensed system represent Bogoliubov quasi-particles, also referred to as \textit{bogolons}. An explicit form of the dispersion law of bogolons depends on the model used to describe the interacting exciton system. In the case of small exciton density, $N_0a_B^2\ll 1$, where $a_B$ is the Bohr radius, an appropriate theoretical model is the Bogoliubov model of weakly-interacting Bose gas. In the framework of this model, the dispersion law of elementary excitations has the form: $\varepsilon_k=\sqrt{\frac{k^2}{2M}\left(\frac{k^2}{2M}+2g_0n_c\right)}$. Here, $n_c$ is exciton density in the condensate, $g_0$ is exciton-exciton interacting constant which can be estimated as $g_0\approx 4\pi e^2d/\epsilon$, and $M$ is the exciton mass. In the long-wavelength limit, $\frac{k^2}{2M}\ll 2g_0n_c$, elementary excitations represent the sound quanta, $\varepsilon_k\approx sk$, where $s=\sqrt{g_0n_c/M}$ is their velocity.

Further, in order to find the response function of the condensate, we use the Gross-Pitaevskii equation:
\begin{gather} \label{eq5}
i\partial_t\Psi(\textbf{r},t)=\left(\frac{(\hbar\hat{\textbf{k}})^2}{2M}-\mu+g_0|\Psi(\textbf{r},t)|^2\right)\Psi(\textbf{r},t)+\\\nonumber
+\Psi(\textbf{r},t)\int d\textbf{r}'V(\textbf{r}-\textbf{r}')\delta n(\textbf{r}',t),
\end{gather}
where the last term describes electrostatic interaction of excitons with the fluctuating electron density. We will treat it here as a perturbation, within the linear response approach. The wave function of the condensed particles, $\Psi(\textbf{r},t)$, can be split into the stationary uniform part and the perturbed contribution: $\Psi(\textbf{r},t)=\sqrt{n_c}+\psi(\textbf{r},t)$. Further, the response function of the condensate excitons, $P_{k\omega}$, can be defined as
\begin{equation}
\label{EqExFluct}
\delta N_{k\omega}=P_{k\omega}V_{k}\delta n_{k\omega},
\end{equation}
where $\delta N_{k\omega}=\sqrt{n_c}(\psi^{\ast}(\textbf{r},t)+\psi(\textbf{r},t))$ is a perturbation of the condensate particle density. Then, the linearization of (\ref{eq5}) gives:
\begin{eqnarray} \label{eq6}
P_{k\omega}=\frac{n_ck^2/M}{(\omega+i\delta)^2-\varepsilon_k^2}.
\end{eqnarray}
It should be noted, that simple consideration described above disregards the processes of exciton scattering on impurities, developed in~\cite{RefGergelSurisJETP,RefKovalevChaplikJETP2016}.

In our recent work~\cite{RefKovalevSavenkoArXive}, it has been demonstrated that the exciton-impurity scattering results in the finite value of the lifetime of the bogolons, $\varepsilon_k=sk-i\gamma_k$, where
\begin{eqnarray} \label{eq7}
\gamma_k=\frac{1}{\tau_X}(k\xi)^3
\end{eqnarray}
under the assumption $k\xi\ll 1$. Here $\tau_X$ is exciton-impurity scattering time in normal exciton gas state and $\xi=1/2Ms$ is the healing length.
\begin{figure}[!t]
	\includegraphics[height=0.6\linewidth]{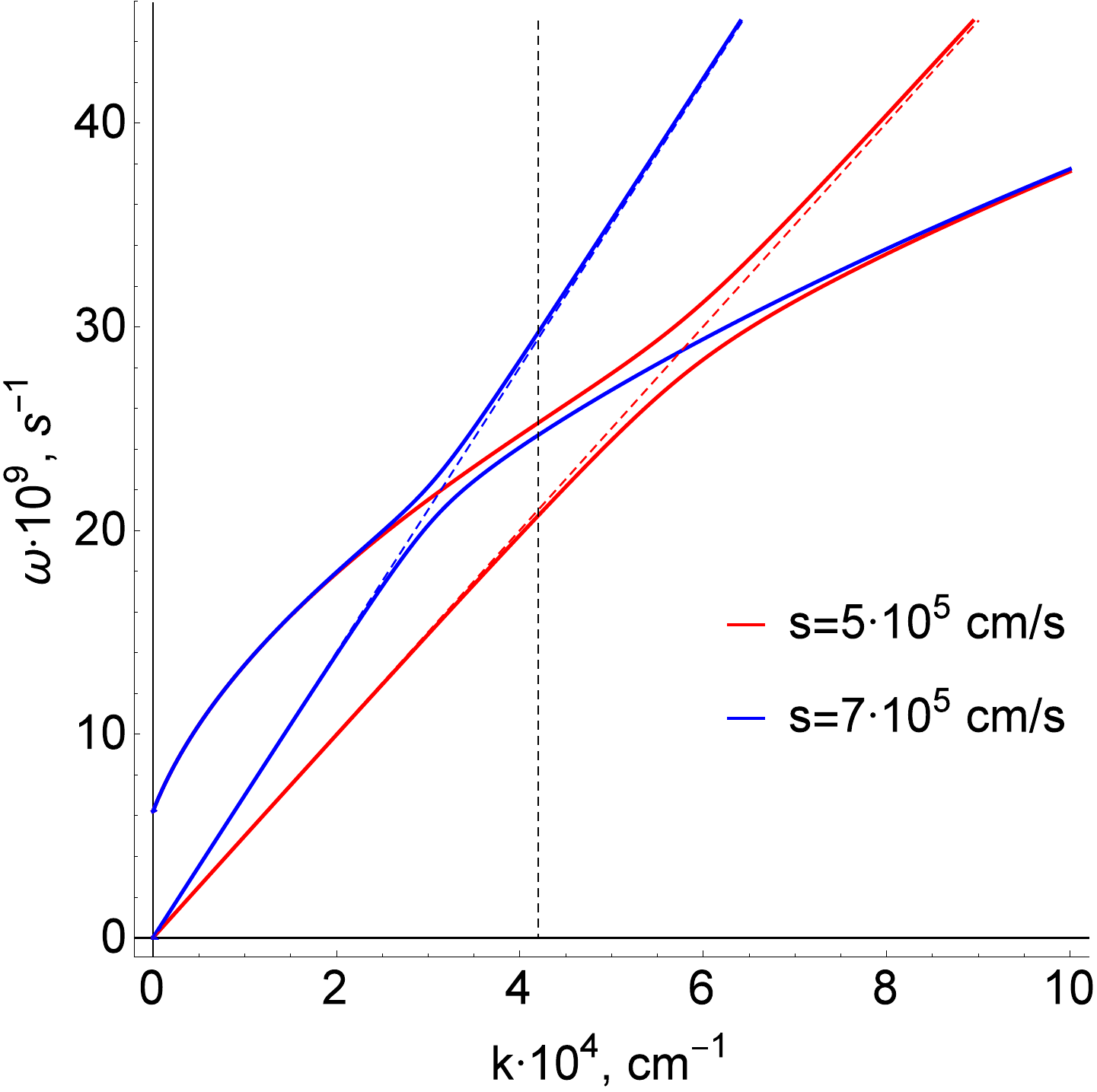}
	\caption{Dispersion curves of the hybrid plasmon-exciton modes at two different values of the Bogoliubov sound velocity. The vertical dotted line stands for $k=4.2\cdot10^4 cm^{-1}$, see main text.}
	\label{Fig2}
\end{figure}
\begin{figure}[!b]
	\includegraphics[height=0.5\linewidth]{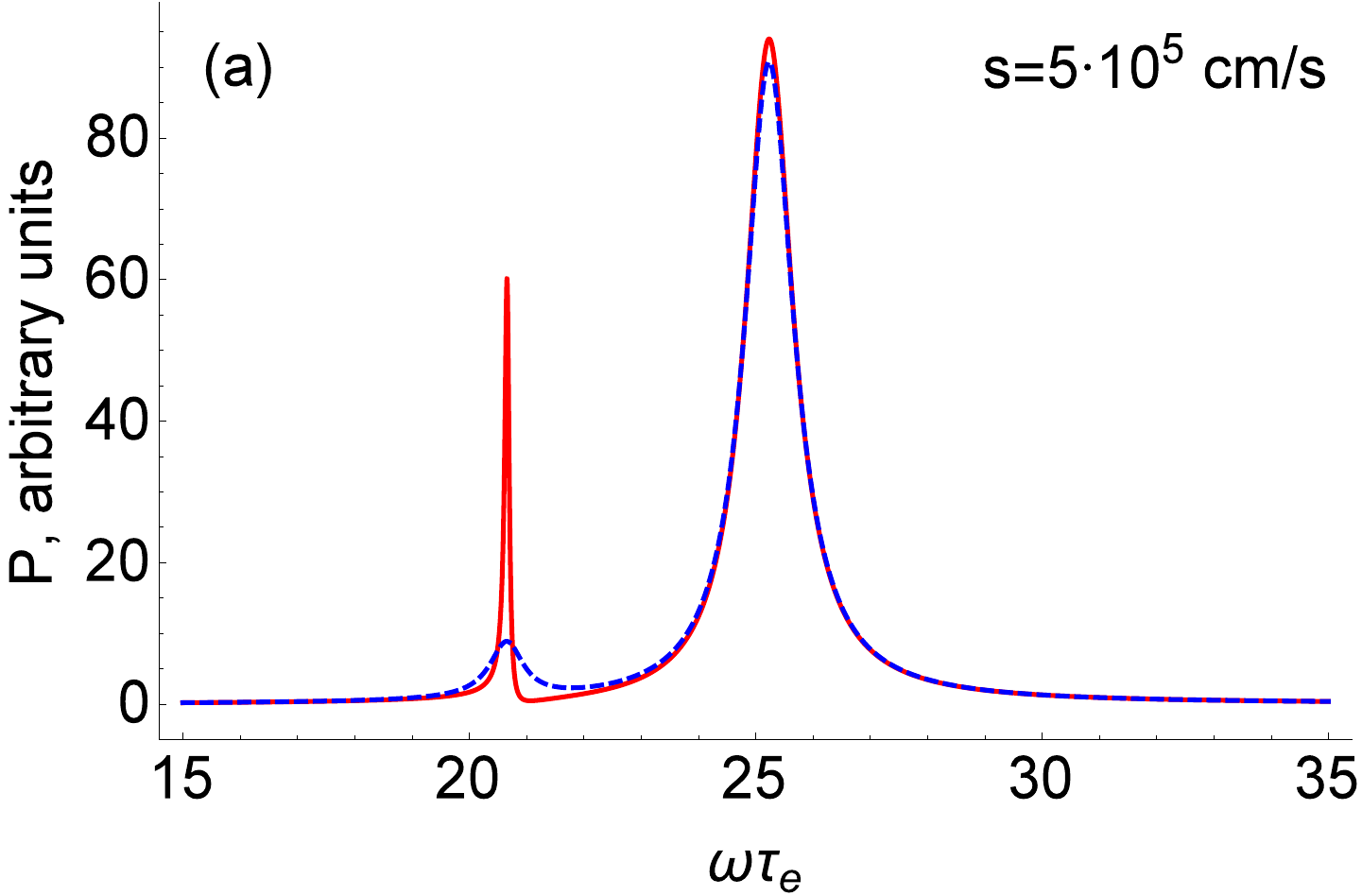}
	\includegraphics[height=0.5\linewidth]{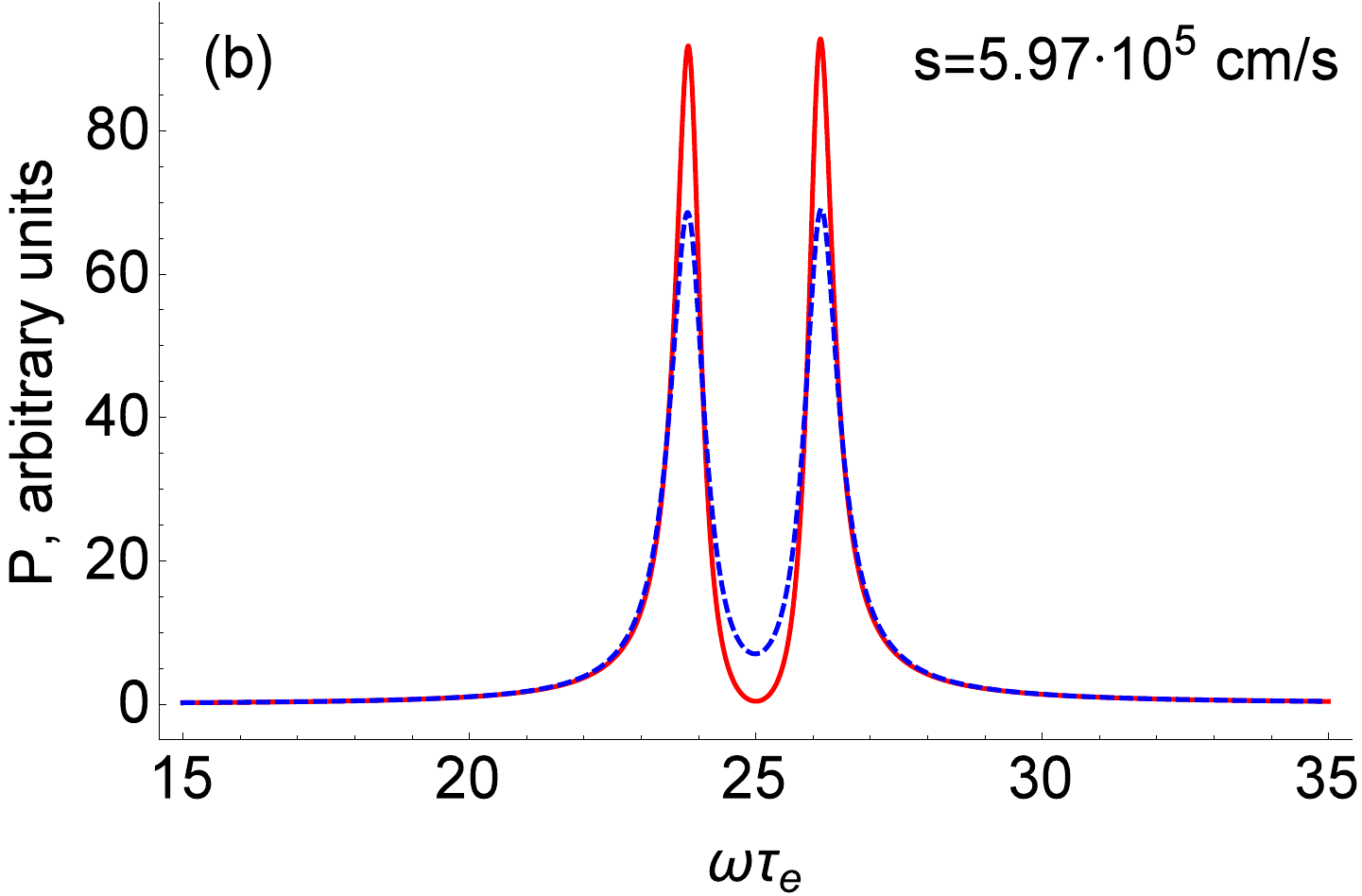}
	\includegraphics[height=0.5\linewidth]{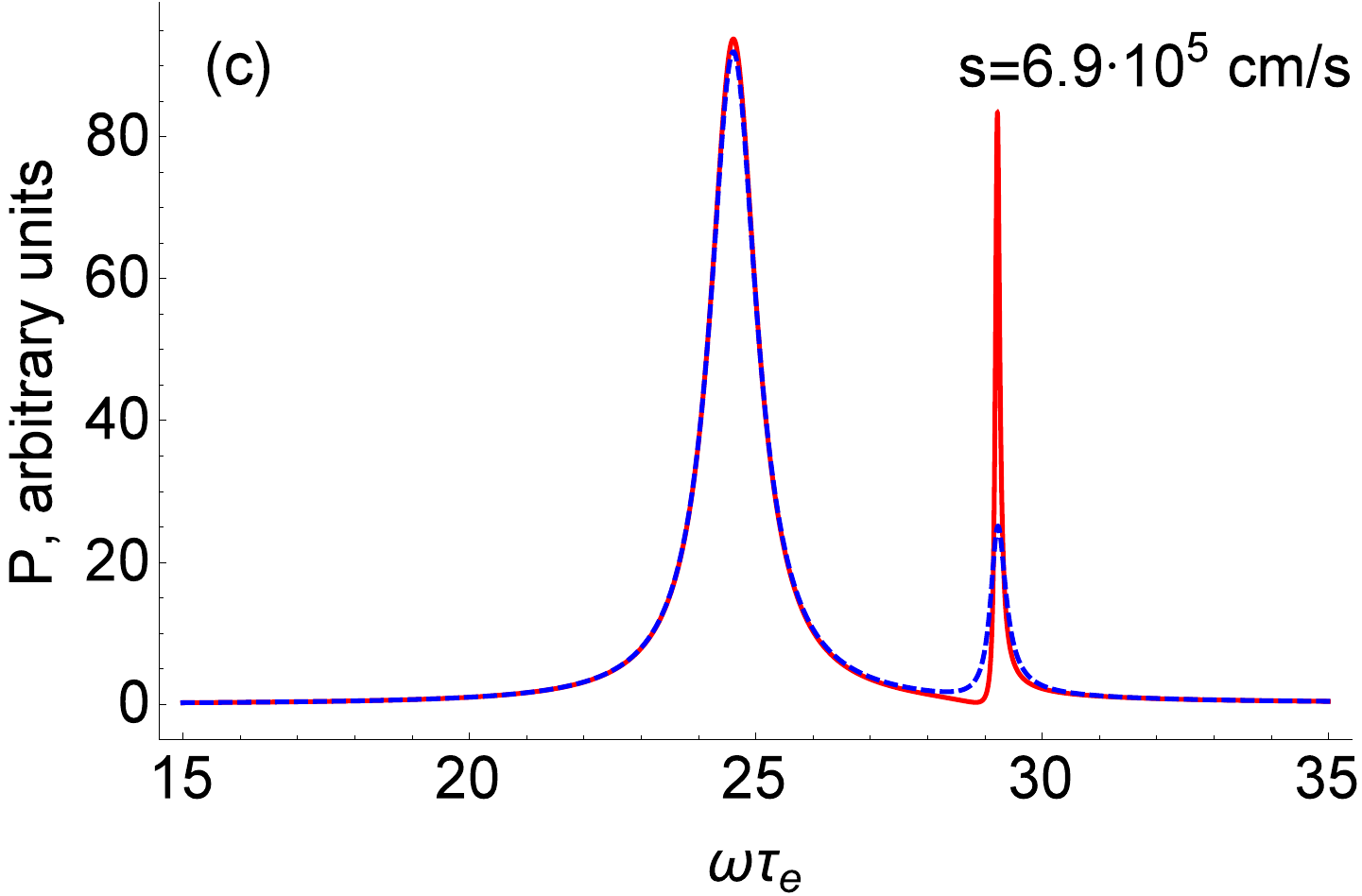}
	\caption{Manifestation of the Fano resonance in the hybrid electron-exciton system presented in Fig.~\ref{FigSystem}. The power absorption (See Eq.~\eqref{eqPowerAbs}) as a function of frequency for the increasing $s$ from (a) to (c) for $\tau_X=10^{-8}$ s (red solid curves) and $ \tau_X=5\cdot10^{-10}$ s (blue dashed curves). The positions of the resonances at the top and bottom panels correspond to the values of excion-plasmon modes taken at $k=4.2\cdot10^4$ $cm^{-1}$ in Fig.~\ref{Fig2}}
	\label{Fig3}
\end{figure}
Thus, the response function, Eq.~(\ref{eq6}), should be modified according to the substitution: $\delta\rightarrow \gamma_k$.

Combining Eq.~\eqref{EqExFluct} together with Eqs.~(\ref{eq1}), (\ref{eq2}), (\ref{eq3}) and defining the renormalised conductivity as $j_{k\omega}=\tilde{\sigma}_{k\omega}E_0$ yields:
\begin{gather}\label{eq?}
\tilde{\sigma}_{k\omega}=\frac{1}{\sigma_B^{-1}+i\frac{k^2}{e^2\omega}\left[U_{k\omega}+V^2_{k\omega}P_{k\omega}\right]},\\\nonumber
\sigma_B=\sigma_0\frac{i(\omega\tau_e+i)}{(\omega\tau_e+i)^2-\omega_c^2\tau_e^2},
\end{gather}
where $\omega_c=eB/mc$ is electron cyclotron frequency, $\tau_e$ is electron relaxation time, and $\sigma_0=e^2n_0\tau_e/m$ is static Drude conductivity of electron gas. Now we can find the EM power absorption as the real part of the conductivity:
\begin{gather}
\label{eqPowerAbs}
P_{k\omega}=\frac{1}{2}E_0^2\,\textmd{Re}\,\tilde{\sigma}_{k\omega}.
\end{gather}

It is obvious, that due to the Coulomb interaction between excitons and electrons, new hybrid modes appears, describing the joint oscillation of exciton and electrons densities. The dispersion law of these new modes can be easily determined by the poles of the renormalized conductivity, $\tilde{\sigma}^{-1}_{k\omega}=0$. In the limit of large relaxation times, $\tau_e,\tau_X\rightarrow\infty$, the renormalized conductivity poles satisfy the equation:
\begin{gather}
\label{dispeq}
\left(\omega^2-\omega_p^2\right)\left(\omega^2-\varepsilon_k^2\right)-\beta_k^2=0,
\end{gather}
where magnetoplasmon dispersion is given by $\omega_p=\sqrt{\omega_c^2+\omega_k^2}$ and $\omega_k^2=2\pi e^2n_0k/\epsilon m$ is a plasmon dispersion of the 2D electron gas and the coupling parameter reads:
\begin{equation}
\nonumber
\beta_k=\omega_k^2\sqrt{\frac{mn_c}{Mn_0}}\left(1-e^{-kd}\right)e^{-ka}.
\end{equation}
Solution of Eq.~\eqref{dispeq} gives the dispersion of two new density oscillation modes:
\begin{gather}
\label{displaws}
\omega_{1,2}^2=\frac{1}{2}\left(\omega_p^2+\varepsilon_k^2\right)\pm\frac{1}{2}\sqrt{\left(\omega_p^2-\varepsilon_k^2\right)^2+4\beta^2_k}.
\end{gather}
Figure~\ref{Fig2} demonstrates the dispersion law of these new modes calculated at two different values of bogolon phase velocity, $s$.


{\it Results and Discussion.---}
Let us investigate the properties of the power absorption function. In the calculations we use the following parameters: $d=10^{-6}$ cm, $a=10^{-5}$ cm, $k=4.2\cdot10^{4}$ cm$^{-1}$, and $\sigma_0=3.2\cdot 10^{-5}$ cm$/$s.

Figure~\ref{Fig3} demonstrates the power absorption spectrum of the system with emergence of the Fano resonance. Increasing the velocity of the Bogoliubov excitations, $s$, we shift the Fano resonance in frequency. In particular, Figs.~\ref{Fig3}a,c correspond to the celebrated asymmetric Fano profiles, whereas Fig.~\ref{Fig3}b corresponds to the symmetric configuration. It should also be mentioned, that the mutual disposition of the peaks and the dip of our asymmetric resonance is changed from Figs.~\ref{Fig3}a to Figs.~\ref{Fig3}c.

\begin{figure}[!t]
	\includegraphics[height=0.5\linewidth]{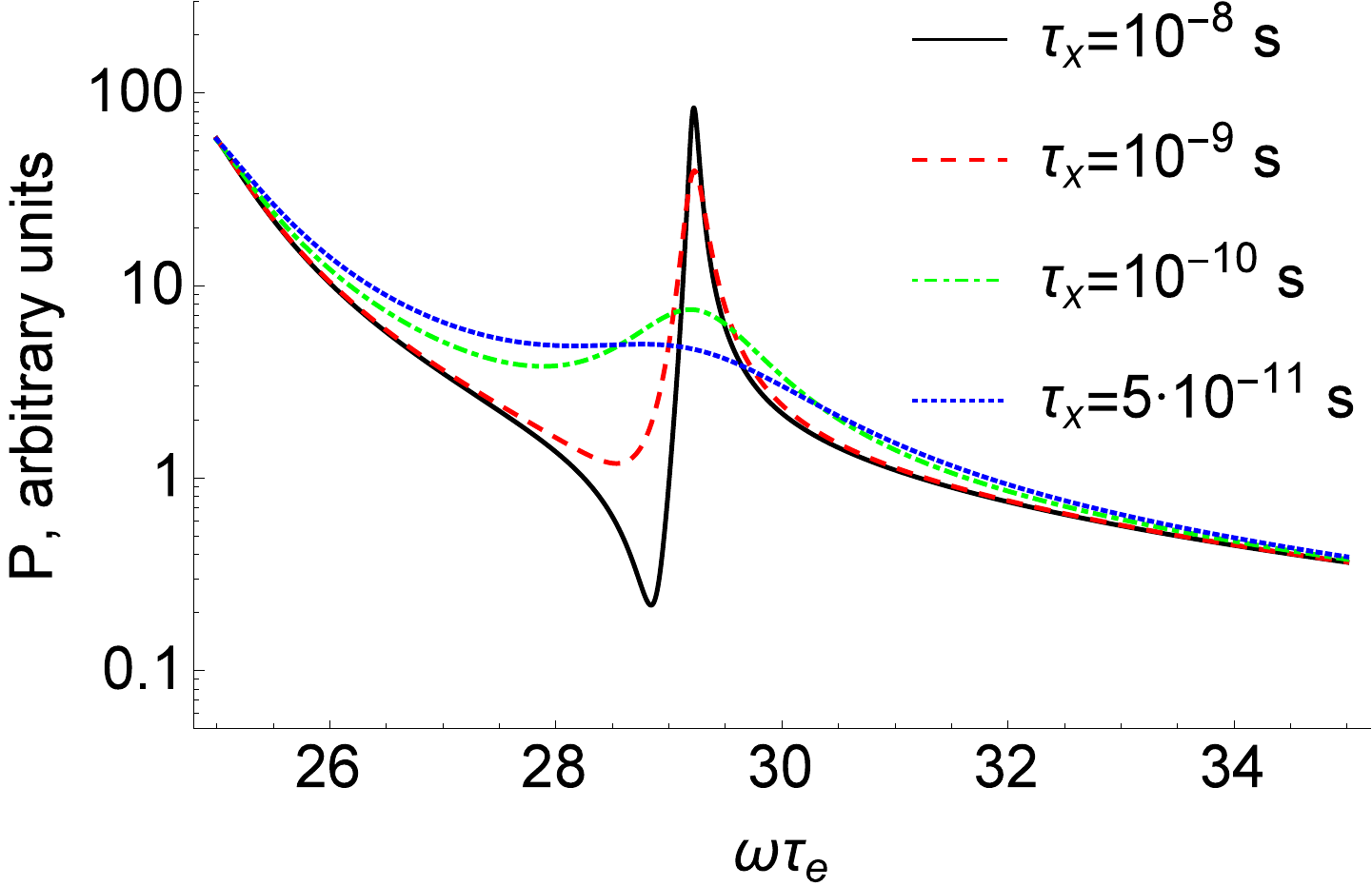}
	\caption{The power absorption spectrum calculated using Eq.~\eqref{eqPowerAbs} for different values of $\tau_X$ in the range of high $\omega\tau_e$ ($s=6.9\cdot 10^5$ cm/s is the same as in Fig.~\ref{Fig3}c).}
	\label{Fig4}
\end{figure}

Figure~\ref{Fig4} shows the power absorption spectrum in the log scale for a fixed value of $s=6.9\cdot 10^5$ cm/s
(corresponding to Fig.~\ref{Fig3}c) and different values of the impurity-mediated scattering rates.
The frequency range is restricted to the high-$\omega$ region corresponding to the second peak in Fig.~\ref{Fig3}c.
It is clearly seen that with the increase of the excitonic impurity-mediated lifetime
(which corresponds to the decrease of the influence of the impurity-assisted scattering),
the resonant peak becomes more Fano-like and we observe the emergence of a dip in the power absorption spectrum around $\omega\tau_e=29$.

\begin{figure}[!t]
	\includegraphics[height=0.5\linewidth]{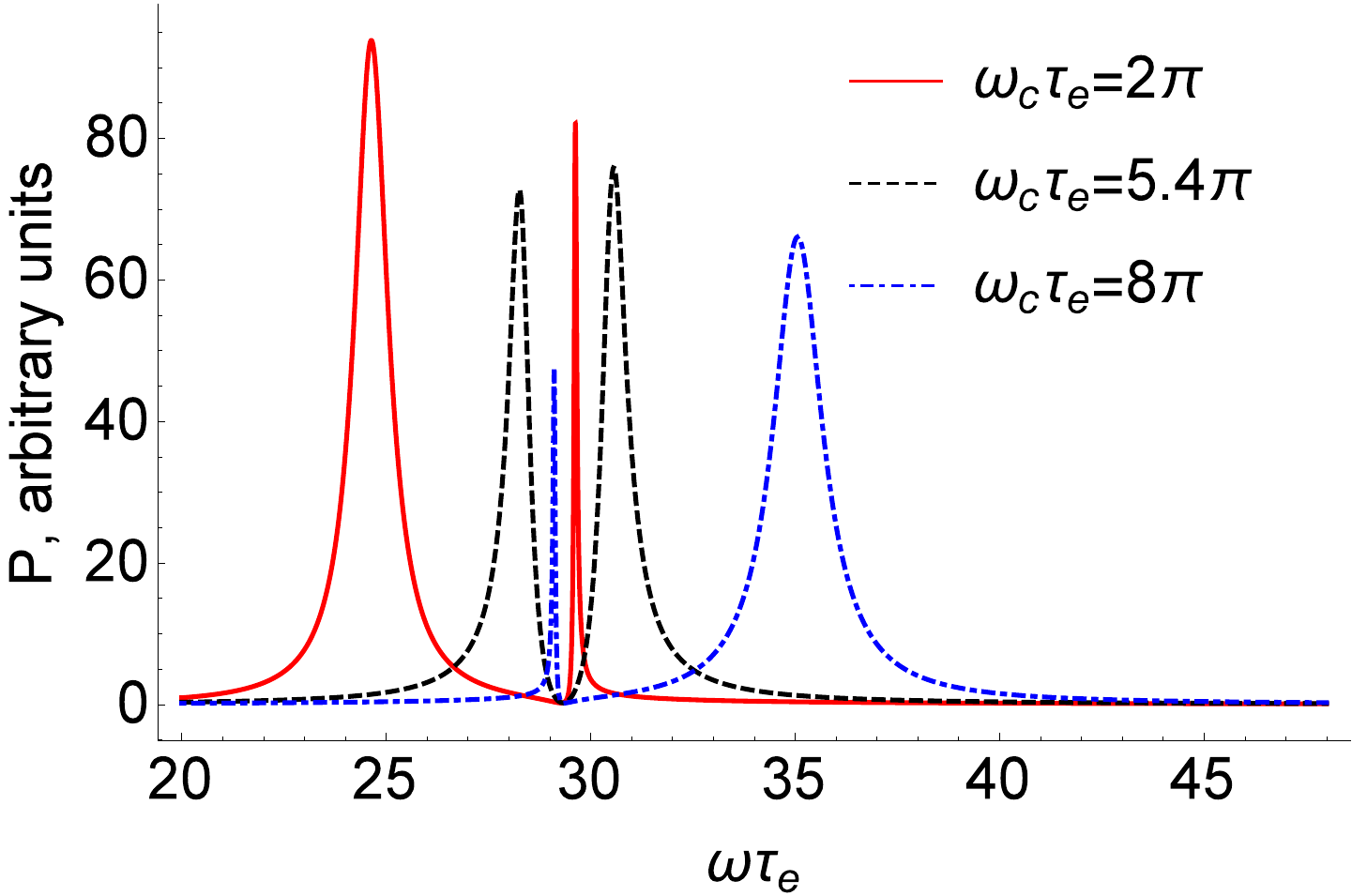}
	\caption{Magnetic field effect. Power absorption spectrum at $s=7\cdot10^5$ cm/s and different values of magnetic field, $\omega_c\tau_e$. Compare the alteration of the spectra caused by the influence of the magnetic field and the increase of the bogolon velocity (see Fig.~\ref{Fig3})}
	\label{Fig5}
\end{figure}

Figure~\ref{Fig5} explains the dependence of the power absorption coefficient on the magnitude of the external applied magnetic field presented in units of $\omega_c\tau_e$. The increase of the magnetic field
($\omega_c\tau_e$) gives a similar effect as the increase of $s$ (see also Fig.~\ref{Fig2}).
However, it should be noticed, that the position of the magnetoplasmon resonance is determined by
the magnetic field strength, whereas the Fano-type resonance coming from the exciton
subsystem is determined by the Bogliubov sound velocity, $s$ (compare
Figs.~\ref{Fig3} and~\ref{Fig5}). The latter, in turn, is determined by the exciton condensate density since $s\sim\sqrt{n_c}$. These dependencies can, for instance, be used to experimentally control the spectral properties of the system under consideration.

It is also important to mention, that if it happens that the plasmon modes are strongly damped, then their dispersion branches turn into the electron-hole continuum. Thus, in order to observe the predicted effect, it is crucial that the following inequality holds: $s>v_F$, where $v_F$ is the Fermi velocity of the electron gas.
This inequality can always be satisfied at typical exciton densities $10^{10}\div10^{12}$~$cm^{-2}$ used in modern experimental DQW structures.
%
%
%

%
%
%


{\it Conclusions.---}
We have developed a microscopic theory of magnetoplasmon resonance in a hybrid structure consisting of interacting electron and dipolar exciton gases. We show that the Coulomb interaction between them results in hybridisation of the density waves of both the subsystems with the appearance of new double-peak dispersion. We attribute the emergence of this hybrid mode to the cyclotron resonance effect. It has also been shown that the power absorption spectrum demonstrates several interesting features, in particular, the Fano-type shape of the resonance in relatively pure samples. We expect that these new modes can be easily found experimentally by well developed techniques, used in study of plasmons in low-dimensional nanostructures.


We thank Profs. A. Chaplik, A. Miroshnichenko and S. Flach for useful discussions and critical reading of the manuscript.
V.M.K and M.V.B. acknowledge the support from RFBR grant $\#16-02-00565a$.
I.G.S. acknowledges support of the Project Code (IBS-R024-D1), Australian Research Council Discovery Projects funding scheme (Project No. DE160100167), President of Russian Federation (Project No. MK-5903.2016.2). I.G.S. and M.V.B. also thank the Dynasty Foundation.



\begin{thebibliography}{30}

\bibitem{RefImamogluPRB2016} O. Cotlet, S. Zeytinoglu, M. Sigrist, E. Demler, and A. Imamoglu, \textit{Phys. Rev. B} \textbf{93}, 054510 (2016).

\bibitem{RefFesh1} S. Inouye, J. Goldwin, M. L. Olsen, C. Ticknor, J. L. Bohn, and D. S. Jin, \textit{Phys. Rev. Lett.} \textbf{93}, 183201 (2004).

\bibitem{RefFesh2} E. Wille, F. M. Spiegelhalder, G. Kerner, D. Naik, A. Trenkwalder, G. Hendl, F. Schreck, R. Grimm, T. G. Tiecke, J. T. M. Walraven, S. J. J. M. F. Kokkelmans, E. Tiesinga, and P. S. Julienne, \textit{Phys. Rev. Lett.} \textbf{100}, 053201 (2008).

\bibitem{RefFesh3} F. M. Marchetti, C. J. M. Mathy, D. A. Huse, and M. M. Parish, \textit{Phys. Rev. B} \textbf{78}, 134517 (2008).

\bibitem{RefFesh4} D. A. Brue and J. M. Hutson, \textit{Phys. Rev. Lett.} \textbf{108}, 043201 (2012).

\bibitem{RefPhase1} D.-W. Wang, M. D. Lukin, and E. Demler, \textit{Phys. Rev. A} \textbf{72}, 051604(R) (2005).

\bibitem{RefPhase2} H. P. B\"uchler and G. Blatter, \textit{Phys. Rev. Lett.} \textbf{91}, 130404 (2003).

\bibitem{RefShelykhPRL2010} F. P. Laussy, A. V. Kavokin, and I. A. Shelykh, \textit{Phys. Rev. Lett.} \textbf{104}, 106402 (2010)

\bibitem{RefButov} L. V. Butov, \textit{Sol. State Comm.} 127, 89 (2003); \textit{J. Phys.: Cond. Matt} \textbf{16}, R1577 (2004); \textit{J. Phys.: Cond. Matt.} \textbf{19}, 295202 (2007).

\bibitem{RefTimofeev} V. B. Timofeev, A. V. Gorbunov, \textit{Physica Stat. Solidi C, Current Topics in Solid State Physics} \textbf{5}(7) pp. 2379?2386 (2008) and JETP \textbf{84}(6) 390?396 (2006).

\bibitem{RefShelykhPRL1051404022010} I. A. Shelykh, T. Taylor, and A. V. Kavokin, \textit{Phys. Rev. Lett.} \textbf{105}, 140402 (2010).

\bibitem{RefMatuszewskiPRL1080604012012} M. Matuszewski, T. Taylor, and A. V. Kavokin, \textit{Phys. Rev. Lett.} \textbf{108}, 060401 (2012).

\bibitem{RefFano1} A. E. Miroshnichenko, S. F. Mingaleev, S. Flach, and Yu. S. Kivshar, \textit{Phys. Rev. E} \textbf{71}, 036626 (2005).

\bibitem{RefFano2} U. Siegner, M. -A. Mycek, S. Glutsch, and D. S. Chemla, \textit{Phys. Rev. Lett.} \textbf{74}, 470 (1995).

\bibitem{RefMiroshnichenkoFanoReview} A. E. Miroshnichenko, S. Flach and Yu. S. Kivshar, \textit{Rev. Mod. Phys.} \textbf{82}, 2257 (2010).

\bibitem{RefGergelSurisJETP} V. A. Gergel, R. F. Kazarinov, and R. A. Suris, \textit{Sov. Phys. JETP} \textbf{31}(2) pp. 367-373 (1970).

\bibitem{RefKovalevChaplikJETP2016} V. M. Kovalev and A. V. Chaplik, \textit{J. Exp. Theor. Phys.} \textbf{122} 499 (2016).

\bibitem{RefKovalevSavenkoArXive} V. M. Kovalev and I. G. Savenko, \textit{arXiv:1609.06411} (2016).





\bibitem{RefKovalevSavenkoIorsh} V. M. Kovalev, I. G. Savenko, I. V. Iorsh,
\textit{J. of Phys.: Cond. Mat.} \textbf{28}(10) 105301 (2016).

\end{thebibliography}
\end{document}